\documentclass[12pt]{article}

\newtheorem{theorem}{{\bf \sc Theorem}}

\def\eproof{\hbox{\hskip3pt\vrule width4pt height8pt depth1.5pt}}
\def\Re{I\!\!R}

\begin{document}

\title{Average Distance, Diameter, and Clustering in Social Networks with
Homophily \thanks{%
Financial support from the
NSF under grant SES--0647867 is gratefully acknowledged. I thank Ben Golub for helpful conversations.
Written for the Workshop on Internet and Network Economics (WINE 2008), in the Lecture Series on Computer Science, Springer-Verlag.}}
\author{Matthew
O. Jackson\thanks{%
Department of Economics, Stanford University and the Santa Fe Institute.
Email: jacksonm@stanford.edu, http://www.stanford.edu/$\sim$jacksonm/} }
\date{September 24, 2008 
}
\maketitle

\begin{abstract}
I examine a random network model where nodes are categorized by type and linking probabilities can differ across types.  I show that as
homophily increases (so that the probability to link to other nodes of the same type increases and the probability of linking
to nodes of some other types decreases) the average distance and diameter of the network are unchanged, while the average clustering in the network increases.  
\medskip


\medskip

\end{abstract}

\section{Introduction}

Communication advances and the social networking via the Internet have made it much easier for individuals to locate others with similar backgrounds and tastes.   This can affect the formation of social networks.  
How do such changes in the ability of individuals to locate other similar individuals affect social network structure? 
Answering this question requires having models of how homophily, the tendency of nodes to be linked to other nodes with similar characteristics, 
affects social network structure.   Homophily is a well-studied and prevalent phenomenon that is observed across all sorts of applications and attributes including ethnicity, age, religion, gender, education level, profession, political affiliation, and other attributes (e.g., see Lazarsfeld and Merton (1954), Blau (1977), Blalock (1982), Marsden (1987, 1988), among others, or the survey by McPherson, Cook and Smith-Lovin (2001)).
Despite the extensive empirical research on homophily, there is little that is known about how homophily changes a network's basic characteristics, such as the average distance between nodes, diameter, and clustering.     

This paper examines the following questions.    Given is a society of nodes that are partitioned into a number of different groups where nodes within a group are of the same ``type'' and nodes in different groups are of different types.  A network formation process is examined that can embody various forms of homophily: 
the probability of links between pairs of nodes can depend on their respective types.  
Holding the degree distribution constant, how does such a network that is formed with substantial homophily compare to a network formed when types are ignored?   
One conjecture is that as homophily increases so that the probability of links among nodes of similar types increases and the probability of links across less similar types falls, the average distance and diameter of the network will increase since the density of links across different types of nodes will be falling. 
This conjecture turns out to be false.  Even as the probability of links across types falls, the  average distance and diameter are not changed even in some extreme cases where the relative probability a link between nodes of the same type is arbitrarily more likely than a link among nodes of different types, provided some non-vanishing fraction of a node's links are still formed to nodes of other types. 
In contrast, homophily can have a significant impact on clustering.  It is shown that substantial homophily can lead to nontrivial clustering, while a process with the same expected degrees but no homophily exhibits no clustering.

\section{A Model of Network Formation with General forms of Homophily and Degree Sequences}

A network $G=(N,g)$ is a graph that consists of a set $N=\{1,\ldots,n\}$ of a finite number $n$ of nodes along with a list of edges, $g$,\footnote{Formally, $g\subset 2^N$ such that each element in $g$ has cardinality 2.} which are the pairs of nodes that linked to each other.  

Given that the network might not be connected, I follow Chung and Lu (2002) in defining average distance in the network to be the average across pairs of path-connected nodes. 
In particular, let $\ell_g(i,j)$ be the number of links in the shortest path connecting nodes $i$ and $j$ if there is such a path, and let $\ell_g(i,j)$ be infinity if there is no path between $i$ and $j$ in $g$.\footnote{Standard definitions, such as path, are omitted.  See Jackson (2008) for such definitions.}  Thus, the average distance in the network is defined as\footnote{Self-loops are allowed here, and so under these definitions if there is a self-loop then a node is a distance of 1 away from itself.  This is irrelevant to the results and simply for convenience.  It is easily seen that the results are the same if self-loops are ignored or if self-distance is set to 0.  If there are no links in the network, the AD expression is 0/0 which can be set to take any value.} 
\[
AD(g)=\frac{\sum_{\{i,j\}: \ell_g(i,j)\neq \infty} \ell_g(i,j)}{ |\{ \{i,j\} : \ell_g(i,j)\neq \infty\}|}.
\]
The diameter of the network is $diam(g)=\max_{\{i,j\}: \ell_g(i,j)\neq \infty} \ell_g(i,j)$.

For the network formation processes considered here, the largest component contains all but at most a vanishing fraction of nodes and so these definitions are effectively the same whether we defined them as above, or just work with the largest component of $g$ which is either the whole network or almost all of it.  

The clustering of a node $i$ with degree of at least 2 is
\[
CL_i(g)=\frac{|\{\{j,j'\} : i\neq j\neq j'\neq i; \{i,j\}\in g, \{i,j'\}\in g, \{j,j'\}\in g \}|}{|\{\{j,j'\} : i\neq j\neq j'\neq i; \{i,j\}\in g, \{i,j'\}\in g \}| }.
\]
The average clustering is the average of $CL_i$ across nodes $i$ that have degree at least 2.\footnote{Set clustering to 0 if there are no such nodes.}

\subsection{A General Random Network Model with Homophily}

The following model is a generalization of the random network model from Chung and Lu (2002) to allow nodes to be of different types and to allow heterogeneous probabilities of linking across different types.

A set of nodes $N=\{1,\ldots,n\}$ is partitioned into $K$ groups or types $N_1,\ldots, N_K$.   This partition captures the characteristics of the nodes, so that all nodes with the same characteristics are in the same group $N_k$.  Depending on the application a type might embody 
ethnicity, gender, age, education, profession, etc. in a social setting, or might involve characteristics of a business in a market network, or might involve some physical characteristics of a node in a physical network.

Also given is a degree sequence $\{d_1,\ldots,d_n\}$ which indicates the expected degree or number of connections of each node.
Let 
\[
D=\sum_i d_i 
\]
and
\[
\widetilde{d}=\sum_i d_i^2/ D. 
\]
Note that if $d_i=d$ for all $i$, then $\widetilde{d}=d$.

Let $D_k=\sum_{i\in N_k} d_i$ be the total degree of all nodes of type $k$.

A random network is formed according to the following process.    
For each pair of types $k$ and $k'$ there is a parameter $h_{kk'}\geq 0$.  This parameter captures the relative proclivity of groups $k$ and $k'$ to link to each other.
The parameters satisfy
$\sum_{k'} D_{k'} h_{kk'} =  D$
for each $k$.
A link between nodes $i$ in group $k$ and $j$ in group $k'$ is formed with probability 
\[
h_{kk'} d_i d_j/D.
\]
Conditions defined below ensure that this expression does not exceed 1.

In the case where $h_{kk}>h_{kk'}$ for all $k$ and $k'\neq k$, then there is homophily, so that nodes are relatively more likely to form their links to their own types than to other types.
If $h_{kk'}=1$ for all $k$ and $k'$ then types are irrelevant and the model reduces to the usual Chung and Lu model.  Otherwise, this allows for different patterns of linkings between
different types.
If $d_i=d$ for all $i$, then this is a generalization of Erd\"os-Renyi random graphs where links are type-dependent.\footnote{Note, however, that this process allows for self-loops $i$ may connect to $i$, although the probability of this for any node $i$ vanishes as $n$ grows provided $d_i^2/D$ vanishes.}   
More generally, the degree distribution could vary across nodes, and 
power-law networks are the special case where the frequency distribution of $\{d_1,\ldots,d_n\}$ has a power distribution where the frequency of degree $d$ is of the form $c d^{-\gamma}$ for some range of $d$.

An interesting case is where types have some social or spatial geography and type $k$ can be represented as a vector $x_k\in \Re^m$ for some $m$ and then $h_{kk'}$ is decreasing
in the distance between $k$ and $k'$; for example of the form $c-f(|x_k-x_{k'}|)$ where $c$ is a constant and $f$ is an increasing function.  One can also 
consider some hierarchy among the $k$'s with the relative probabilities depending on the hierarchy (e.g., see Clauset, Moore and Newman (2008)).
Another case of interest is where types have a given probability of forming links to their own type and a different probability of forming links all other types (e.g., see Copic, Jackson and Kirman (2005) and Currarini, Jackson and Pin (2007)). 
 
\subsection{Admissible Models}

The main results consider a growing sequence of network formation models, and so all parameters are indexed by $n$, the number of nodes.
The results use some restrictions on variation in expected degrees across nodes and a minimum bound on the proclivity to link across groups.   A sequence of network formation processes is said to be {\sl admissible} if the following conditions are satisfied.

First, there exists $h>0$ such that $h_{kk'}(n)>h$ for all $k$ and $k'$ for all large enough $n$.
This condition does {\sl not} require that nodes of different types have a probability of linking that is bounded below, as a node's degree could be a fixed number independent of $n$. 
This lower bound simply implies that any given node spreads some of its links on types other than its own type.   This still allows for extreme homophily, as it can still be that $h_{kk}(n)\rightarrow \infty$ and that
the probability of links with own type is becoming infinitely more likely than links with some other types.

Second, the degree sequence satisfies the following:
\begin{itemize}
\item
there exists $\varepsilon>0$ such that
$\widetilde{d}(n)\geq (1+\varepsilon) \log(n)$ for large enough $n$ and $\log\left(\widetilde{d}(n)\right)/\log(n) \rightarrow 0$
\item
there exists $c >0$ such that $ h c>1$, and $M>0$, such that 
$d_i(n)\leq M\widetilde{d}(n)$ for all $i$ and $n$, and $d_i(n)\geq c$ for all but $o(n)$ 
nodes.\footnote{Here, $h$ is as defined in the restrictions on proclivity to link across types. 
These conditions ensure that the degree sequence satisfies (i) and (ii) in Chung and Lu (2000).  They also guarantee 
(iii) setting $U=N$ and noting that $\widetilde{d}(n)\leq M^2 D(n)/n$.}
\end{itemize}
The first restriction is that the second-order average degree is growing with $n$, but more slowly than $n$. It grows at a rate fast enough that the giant component includes a fraction approaching 1 of the nodes.   The second requires that no node have an expected degree that explodes relative to the average expected degree and that all but a vanishing fraction of nodes have a lower bound on expected degree that is larger than 1.

\section{Diameter and Average Distance in the Model}

Let $AD(n,\mathbf{d}(n),\mathbf{h}(n))$ and $diam(n,\mathbf{d}(n),\mathbf{h}(n))$ be the average distance and diameter, respectively, of a graph randomly drawn according to the process above with  $n$ nodes, degree sequence $\mathbf{d(n)}=(d_1(n),\ldots,d_n(n))$, and homophily parameters $\mathbf{h}(n)=\left(h_{kk'}(n)\right)_{kk'})$.  This average distance and diameter are random variables for each $n$.  Similarly, let $AD(n,\mathbf{d}(n))$ and $diam(n,\mathbf{d}(n))$ be the average distance and diameter, respectively, of a graph randomly drawn according to the process above with $n$ nodes, degree sequence $\mathbf{d(n)}=(d_1(n),\ldots,d_n(n))$, and without any homophily (so that $h_{kk'}(n)=1$ for all $k$ and $k'$). 

\begin{theorem}
\label{avgdegree}
Consider an admissible sequence of network formation processes $(n,\mathbf{d}(n),\mathbf{h}(n))$.
Asymptotically almost surely:
\begin{itemize}\item  $AD(n,\mathbf{d}(n),\mathbf{h}(n))=\left(1+o(1)\right)\log(n)/\log(\widetilde{d}(n))$, and 
so $\frac{AD(n,\mathbf{d}(n),\mathbf{h}(n))}{AD(n,\mathbf{d}(n))}\rightarrow 1$,
\item $diam(n,\mathbf{d}(n),\mathbf{h}(n))=\Theta\left(\log(n)/\log(\widetilde{d}(n))\right)$
and so $diam(n,\mathbf{d}(n),\mathbf{h}(n))=\Theta\left(diam(n,\mathbf{d}(n))\right)$.
\end{itemize}
\end{theorem}

Thus, the average distance and diameter of the admissible processes are not affected by homophily.  Even though there can be an arbitrarily increased density of links within types, and substantial 
decrease in the density of links across types, this does not impact average distance or the diameter in the network.  
In order for homophily to affect these aspects of the network, one would have to have the density of links across most types decrease at a level which vanishes relative to overall degree.
That is, suppose instead that nodes are grouped into evenly sized groups (up to integer constraints) so that $h_{kk'}(n)\leq f(n)$ for all $k$ and $k'$ with $k'\neq k$ for some $f(n)$ such that $f(n)n \widetilde{d}(n)/K(n)$ is bounded above and where $K(n)/n $ is bounded away from 0.
Then, it is easy to check that,\footnote{A lower bound on the average distance is that of a graph where all nodes of 
a given type are agglomerated to become a single node.  There are $K(n)$ nodes in this graph and each of these type-nodes has degree of at most $\widetilde{d}Mf(n)n/K(n)$ which is bounded above
by some $C$.  
The average distance is at least order $\log(K(n))/\log(C)$ which is proportional to $\log(n)$, provided this network has a giant component containing all but at most a vanishing fraction of nodes.
The average distance could only be smaller than this if the connectivity across types drops so low so that the network fragments to smaller components.}  almost surely,   $\frac{AD(n,\mathbf{d}(n),\mathbf{h}(n))}{AD(n,\mathbf{d}(n))}\rightarrow \infty$ 
and so $\frac{diam(n,\mathbf{d}(n),\mathbf{h}(n))}{diam(n,\mathbf{d}(n))}\rightarrow \infty$.

\bigskip

\noindent{\bf Proof of Theorem \ref{avgdegree}:}
Consider a network formation process such that each 
node has expected degree $hd_i$ and $h_{kk'} =1$ for all $kk'$.   This is the process 
$(n,h\mathbf{d}(n))$, and the process $(n,\mathbf{h}(n),\mathbf{d}(n))$ is equivalent to a first running
the process $(n,h\mathbf{d}(n))$ and then adding some additional links.  
Under the admissibility requirement here, $(n,h\mathbf{d}(n))$ is admissible and specially admissible under the definitions of Chung and Lu (2002).
By Lemma 5 in Chung and Lu (2002), almost surely the largest component of a random graph under the process $(n,h\mathbf{d}(n))$
contains all but at most $o(n)$ of the nodes.  By Theorems 1 and 2 in Chung and Lu (2002) the average distance and diameter of this process are almost surely 
$$\left(1+o(1)\right)\log(n)/\log(h\widetilde{d}(n)) = \left(1+o(1)\right)\log(n)/\log(\widetilde{d}(n)),$$ and $$\Theta\left(\log(n)/\log(h\widetilde{d}(n))\right)= \Theta\left(\log(n)/\log(\widetilde{d}(n))\right),$$ respectively.
Since the process  $(n,\mathbf{h}(n),\mathbf{d}(n))$ is equivalent to a first running
the process $(n,h\mathbf{d}(n))$ and then adding some additional links, it then follows directly that a random graph generated in this way
contains all but at most $o(n)$ of the nodes and has average distance and diameter of this process are almost surely bounded above by
$ \left(1+o(1)\right)\log(n)/\log(\widetilde{d}(n))$, and some factor times $\log(n)/\log(\widetilde{d}(n))$, respectively.

Next, let us show that these are also lower bounds.
First, consider a network where all nodes have degree no more than
$M'\widetilde{d}(n)$.   
Consider any node $i$.  
The $T$-th neighborhood of $i$ includes fewer than 
\[
\sum_{t=1}^T \left(M'\widetilde{d}(n)\right)^t =  \frac{\left(M'\widetilde{d}(n)\right)^{t+1} - M'\widetilde{d}(n)}{M'\widetilde{d}(n) -1}
\]
nodes.
Thus, in order to reach all nodes in the largest component from some node in the largest component (which as argued above contains at least $(1-o(n))n$ nodes) it
takes at least  $T(n)= \log((1-o(1))n)/\log\left(M'\widetilde{d}(n)\right)$ steps to reach every other node in
the largest component, almost surely.
Given that $\widetilde{d}(n)\rightarrow\infty$, it follows that 
$T(n)\geq (1-o(1))\log((n)/\log\left(\widetilde{d}(n)\right)$.
The average distance is thus almost surely at least  
\[
\sum_{t=1}^{T(n)} \left(M'\widetilde{d}(n)\right)^t t/n. 
\]
This is at least $(1-o(1))T(n)$, almost surely. 
Thus, the lower bound on average distance is 
$(1-o(1))\log(n)/\log\left(\widetilde{d}(n)\right)$.
The diameter is at least the average distance, and so this is also a lower bound on diameter.  

Let us now show that with a probability going to 1 all nodes have degree of no more than
$2Md$, and then setting $M'=2M$ implies the result.  This probability is at least $\Pi_i \Pr(d_i\leq 2Md)$.
From Fact 1 in Chung and Lu (2002b) it follows that for any given $i$ $\Pr(d_i\leq 2Md)> 1-e^{-Md/3}$ (bounding $E(d_i)$ by $Md$ and setting $\varepsilon$ in their fact to 1).  The overall probability is then 
at most $\left(1-e^{-Md/3}\right)^n$.  Given that $d\geq (1+\varepsilon)\log(n)$, it follows that this expression is
at least $\left(1-\frac{e^{-M\varepsilon \log(n)/3}}{n}\right)^n$ (taking $M/3\geq 1$ without loss of generality in the definition of $M$), which goes to 1 since $e^{-M\varepsilon \log(n)/3}$ goes to 0. So, with a probability of $1-o(1)$ the average distance is
$ \left(1+o(1)\right)\log(n)/\log(\widetilde{d}(n))$.\eproof

\section{Clustering}

Note that in the model with no homophily if $(\max_i d_i(n))^2/D(n)\rightarrow 0$, then the average clustering almost surely tends to 0 simply because the most probable link has a probability that tends to 0. 
In contrast, if groups are relatively small (of the order of average degree) and there is substantial homophily, then average clustering does not vanish.  
Thus, homophilistic networks exhibit the characteristics of the ``small worlds'' discussed by Watts and Strogatz (1998): nontrivial clustering at the same time as having a diameter on the order of a uniformly random graph. 

\begin{theorem}\label{clustering}
Consider a setting such that (i) there is some $m>0$ such that for large enough $n$, $h_{kk}(n) D_k(n)/D(n)>m$ for all $k$, (ii) $\max_i d_i(n)/\max_k |N_k|$ and
$\min_i d_i(n)/\max_i d_i(n)$ are each $\Omega(1)$, and $\max_i d_i(n)>2$.
Asymptotically almost surely, average clustering is $\Omega(1)$.
\end{theorem}

The proof of Theorem \ref{clustering} is straightforward and so only sketched here.  Let $\max_i d_i(n)/\max_k |N_k|>m_1>0$ and
$\min_i d_i(n)/\max_i d_i(n)>m_2>0$ for all large enough $n$.  The probability of a link between any two nodes of the same type is
 at least 
\[
\frac{(m_2 \max_i d_i(n))^2 \min_k h_{kk}}{D} > \frac{(m_2 \max_i d_i(n))^2 m}{\max_k D_k(n)} >  \frac{(m_2 \max_i d_i(n))^2 m}{\max_k |N_k(n)| \max_i d_i(n)} > 
m_2^2 m_1 m >0
\]
for all large enough $n$.
Given that there is a bound $m_3>0$ so that each node has an expectation of forming a fraction of at least $m_3$ of its links within its own group, and the clustering among pairs of
nodes that it is linked to of own type is at least $m_2^2 m_1 m >0$, it follows that the expected clustering of any node
is bounded away from 0 (conditional on it having degree at least 2).  Given that the expected clustering of all nodes are bounded away from 0
(conditional on having at least degree 2), and all nodes have expected degree bounded away from 0 and so a non-vanishing fraction almost surely end up with degree of at least 2, it can then be shown that
the average clustering is almost surely above 0. 

\section{Discussion}

The results here show that substantial homophily and bias in the way that different types of nodes link to each other can be introduced without altering the average distance or diameter of a network.
On one level this might not have been expected, and yet the proof of this is very simple and basically relies on the fact that some rescaling of the degree of a node up to a fixed factor does not alter the asymptotic average distance and diameter of the resulting networks.    
This does not mean that this leaves the properties of the network unchanged, as we have seen with clustering parameters.  Also, as shown in Golub and Jackson (2008), networks with substantial homophily can still behave quite differently, so 
that even though diameter and average distance remain unchanged, the speed of learning can decrease by orders of magnitude and mixing time on such networks can correspondingly increase by orders of magnitude.

\end{document}